\documentclass[12pt,preprint]{aastex}
\usepackage{natbib}
\shorttitle{Distance to the galaxy IC~342}
\shortauthors{{N. A. Tikhonov and O. A. Galazutdinova}}
\begin{document}
\makeatletter

\fontsize{12}{12} \selectfont

\title{Distance to the galaxy IC~342\footnote{Based on observations made with the NASA/ESA Hubble Space Telescope, obtained from the Data Archive at the Space Telescope Science Institute, 
which is operated by the Association of Universities for Research in Astronomy, Inc., under NASA contract NAS5-26555.
 These observations are associated with proposals 10579, 10768}}
\author{{N. A. Tikhonov and O. A. Galazutdinova}}
\affil{Special Astrophysical Observatory, Russian Academy of Sciences,N.Arkhyz, KChR, 369167, Russia}

\begin{abstract}
Based on archival Hubble Space Telescope images, we have performed stellar photometry for a region of the spiral
galaxy IC~342 located in the Milky Way zone. On the constructed Hertzsprung-Russell diagram, we have identified 
the red giant branch and determined the distance modulus for the galaxy by the TRGB (tip of the red giant branch)
method, $(m - M ) = 27.97\pm0.10$, which corresponds to $D = 3.93\pm0.10$ Mpc. The estimated distance puts IC~342 
spatially close to the galaxy Maffei~1 ($D = 4.1$ Mpc) and allows these galaxies to be considered the center of a
single group. 
\end{abstract}

\section*{INTRODUCTION}

Among the nearest groups of galaxies, the IC~342/Maffei group still remains the most puzzling spatial structure.
 Despite intensive studies of individual galaxies that are probable members of this group 
\citep{mcc89,kris95,but99,dav01,dav05,sah02,kar03,fing03}, the question about the population of the group and the spatial
 positions of its individual galaxies remains unsolved. High extinction (light absorption and scattering) in
 gas-dust clouds of our Galaxy prevents the spatial parameters of the group from being determined, because all
 of the probable group members are located in the Milky Way zone. 

An extinction correction based on the papers by Burstein \& Heiles (1982) or Schlegel et al. (1998) yields 
comparatively correct results when objects far from the Milky Way plane are studied. If, however, an object is
 close to the equatorial plane of the Galaxy, then using these papers gives only approximate extinctions inapplicable
 for accurate measurements. It is this uncertainty that creates difficulties in measuring the distances, which entails
 uncertainties in other physical parameters of the IC~342/Maffei group (its luminosity, mass, and dynamical characteristics). 

Since IC~342 is a very massive galaxy (fig.~1), the spatial distribution and kinematics of the dwarf galaxies and
the total mass of the group can be substantially revised if IC~342 is shifted from its spatial position assumed in
 the calculations. In addition, IC~342 is a spiral galaxy with a population resolved into stars and more numerous and 
diverse distance determination methods than is possible for the second massive galaxy of this group, Maffei~1, an 
elliptical galaxy, can be used for it. 

The distance to IC~342 has been determined repeatedly by various methods. The table lists the values obtained.
 We see a great spread in results from 1.5 to 8.0 Mpc, which made it difficult to compile a list of actual galaxies
 in the group. The value of $D = 3.28\pm0.28$ Mpc obtained from the photometry of 20 Cepheids \citep{sah02} has
 been used in recent years. However, since these Cepheids are located in regions with high and variable 
(from region to region) extinction, one might expect the appearance of unsuspected errors in measuring the distance.
 Independent methods of determining the distance to IC~342 can clarify the situation. This is all the more necessary, 
because very large errors in the distances to galaxies determined by the method of Cepheid photometry were detected
 in previous papers of the authors using the Cepheid method \citep{hoe90,hoe94,hoe98}. For comparison, the 
corrected distances for the same galaxies can be found in  Aparicio (1994), Tolstoy et al. (2007) and Aparicio et al. (2000). 

The TRGB (tip of the red giant branch) method \citep{lee93} based on the photometry of red giants, whose accuracy
is as high as or even exceeds that of the Cepheid method \citep{mak06,riz07}, can be an independent
method of determining the distance to IC~342.

\section*{STELLAR PHOTOMETRY} 
The Hubble Space Telescope (HST) archives contain the ACS/WFC images of IC~342 obtained as part of the programs to study
X-ray sources (ID 10579 and ID 10768) on which an old stellar population can be detected. No HST ACS/WFC observations have
been performed specially to determine the distance to IC~342. Since the more numerous WFPC2 images available in the archives
give a photometric limit insufficient for our purposes, we used the WFPC2 images only for the photometry of bright star-forming
regions in IC~342.
 
Unfortunately, the images with the F435W (corresponding to the Johnson B filter) and F606W (with a broader passband 
than that of the Jonson $V$ filter) filters obtained at request ID 10579 or with the F555W (the Johnson V filter)
and F814W (the Kron-Cousins $I$ filter) filters obtained at request ID 10768 are also ill-suited for constructing a
deep Hertzsprung-Russell (or CM) diagram on which the branch of red giants, relatively faint stars compared to young 
supergiants, could be revealed. Therefore, we had to compose an image with the F606W and F814W filters (fig.~2) based 
on the images of the two observing programs using MIDAS to superimpose the images. The intersection of the two superimposed
images (fig.~1) has a sufficient area to construct a representative CM diagram, while old stars (red giants) are clearly
seen on each of the two images. 
 
 The stellar photometry was performed in a standard way using DAOPHOT II in MIDAS \citep{stet94}. The constructed CM diagram
 for the stars of IC~342 is presented in fig.~3. We see wide and populated young supergiant branches, a large set of 
intermediate-age (AGB) stars, and a red giant branch unresolvable into stars on the diagram due to the large number of points.
 The broadening of the young supergiant branches is related mainly to extinction fluctuations both in IC~342 itself and in
 gas-dust clouds of our Galaxy. On the CM diagram (fig.~3), we clearly see that the blue and red supergiant branches are 
shifted relative to their normal positions, because the light reddens as it passes through gas-dust clouds. This shift can
 be used to estimate the average extinction for the stars of IC~342.
  
  \section*{EXTINCTION TOWARD IC~342} 
  
The main difficulty in determining the distance to IC~342 is even not a high extinction but its random variation over
 the body of the galaxy due to the existence of numerous gas-dust clouds both in IC~342 itself and in our Galaxy in the 
path of light propagation from IC~342. 
  
  The extinction fluctuation in the region of stellar photometry can be reduced in two ways: either by reducing the region
 being investigated or by a preliminary search for the region with a minimum extinction fluctuation. Clearly, the necessary
 region must be located outside the regions of intense star formation in IC~342 in which one might expect the presence of gas-dust
 clouds. We used both possibilities. The field found is marked in fig.~2 and the CM diagram for the stars of this field is shown 
in fig.~4. Despite the absence of clear star-forming regions, we nevertheless see the blue supergiant branch on the CM diagram
 for the stars of this field (fig.~4) whose stars are chaotically scattered over the entire field without concentrating into regions.
 This blue supergiant branch has a small width, suggesting that the extinction is constant in this field. The mean color index for
 these blue supergiants in the magnitude range $24 < I < 26$ is $(V -I ) = 0.75$, while the normal color indices for unreddened 
supergiants can change from $(V-I) = 0.0$ to $(V -I) = -0.3$, depending on their age.

For a more accurate extinction determination, we fitted the isochrones with metallicities $Z = 0.008$ and $Z = 0.02$ into the
 CM diagram for the stars of the field being investigated (fig.~4) based on the tables of stellar isochrones 
from Bertelli et al. (1994).
 The isochrones that fit best into the blue supergiant branch of the CM diagram can be easily chosen by varying two parameters, the 
stellar age and the possible extinction (which leads to a reddening of the stars and to a reduction in their apparent brightness).
 To determine the metallicity of young stars, we used red supergiants whose positions on the CM diagram depend strongly on their metallicity.
 It emerged that the red supergiants of the field could not be described by the isochrones with Z = 0.008 for any age, but they were 
satisfactorily described by the isochrones with $Z = 0.02$. The parameters of the isochrones that describe best the CM diagram allowed us to
 determine the extinction for the stars of the field with an  accuracy of $0\fm02$ and, at the same time, to establish the ages and metallicities
 of the red supergiants.
 
 The shift of the blue supergiant branch was found to be 0.82. This means that the reddening for the stars of the field under
 consideration is $E(V-I ) = 0.82 \pm 0.02$, which corresponds to $E(B-V) = 0.60$, and $A_I = 1.16\pm 0.02$. The age of the youngest
 supergiants in the field being investigated turned out to be 20 Myr and their metallicity is $Z = 0.02$, which corresponds to the solar
 metallicity. This probably might be expected, because IC~342 exceeds our Galaxy in luminosity and mass and the stellar metallicity in
 IC~342 must be high. On the CM diagram (fig.~4), we see that older red supergiants with an age of more than 60 Myr are also present 
within the field.

\section*{DETERMINING THE DISTANCE TO IC~342}
 
 To determine the distance by the TRGB method, we used equations from Lee et al. (1993). Using the CM diagram for the field where 
the extinction was determined (fig.~4), we identified the red giants and constructed their luminosity function (fig.~5). The jump
 in the luminosity function and the change in its gradient point to the tip of the red giant branch at $I_{TRGB} = 25.06\pm0.05$. 
The color of the giant branch is $(V-I )_{-3.5} = 2.39$ and $(V-I)_{TRGB} = 2.6$. Using the previously determined extinction for
 the field under consideration, $A_I = 1.16\pm 0.02$, we find the extinction-corrected photometric parameters of the red giant branch:
 $I_{TRGB} = 23.90$, $(V -I)_{-3.5} = 1.57$, and $(V-I)_{TRGB} = 1.77$. Using the paper by Lee et al. (1993), we find the metallicity
 for the red giants of IC~342, $[Fe/H] = -0.99$, and the distance modulus $(m - M) = 27.97\pm0.10$, which corresponds to $D = 3.93\pm0.10$ Mpc.
 The distance to IC~342 that we obtained disagrees, within the error limits, with the distance inferred by  Saha et al. (2002) and
 we see no possibility to somehow smooth over this discrepancy. 
 
\section*{ STAR CLUSTERS IN IC~342}
  
  A visual examination of the HST images and using the CM diagram (fig.~3) allowed us to reveal several star clusters in the region 
being investigated (fig.~2), including three young globular clusters, nine compact clusters, and ten star complexes. The CM diagrams
were constructed for the stars of these objects within the range from $1.\arcsec5$ to $5.\arcsec0$ of their centers. Choosing the most
optimal isochrones for these diagrams from \citep{ber94}, as was done previously when determining the extinction for the
field being investigated, we estimated the stellar metallicities and ages for these objects and the extinction toward them. Figure 6
presents the CM diagrams for the stars of one young globular cluster (fig.~6a) and three small star complexes (figs.~6b,~6c,~6d) that
are marked in fig.~2. The extinction toward these objects varies within the range from $A_I = 1.11$ to $A_I = 1.18$ from object to object,
i.e., comparatively insignificantly. The mean accuracy of determining the extinction is $0\fm02$. This accuracy can deteriorate to 0.04
 for compact objects . and reaches $0\fm01$ for small open star complexes. The isochrones with   $Z = 0.02$ and an age of 25--40 Myr fit
 well into the CM diagram for the young globular cluster with red supergiants (fig.~6a). A check of the isochrones with different 
metallicities ($Z = 0.008$ and $Z = 0.05$) showed that the positions of the stars on the CM diagram for the globular cluster could
 not be described by them. The stellar metallicity determinations for other clusters and complexes are consistent with the metallicity
 $Z = 0.02$ obtained for the young globular cluster. 
  
  As regards the ages of the stellar objects, it should be kept in mind that the brightest stars in each compact cluster that often 
form a close group unresolvable into individual stars remain outside our measurements and are absent on the CM diagrams. Therefore,
 the star clusters can be slightly younger than is indicated on our diagrams in fig.~6. Such close groups of stars are not observed 
in open star complexes and, hence, the stellar ages, metallicities, and extinction are determined most accurately for small
star complexes. Since the field we studied occupies about 3\% of the galaxy's apparent body, one might expect a considerably larger 
number of star clusters of all types in IC~342. 

\section*{THE DWARF GALAXY KK~35} 

This dwarf galaxy appeared in the list of IC~342 satellites in 1997 \citep{huch97}. Using HST WFPC2 images,
 Karachentsev et al. (2003) performed stellar photometry and determined the distance to KK 35 based on the red giants of this galaxy found,
 as was believed by the authors. The measured distance to KK 35 ($D = 3.16 \pm 0.32$) turned out to be close to the distance to IC~342
 itself that the authors assumed to be $D = 3.28 \pm 0.26$ based on the measurements by Saha et al. (2002). An approximate equality of 
the distances to KK~35 and IC~342 seemed to confirm that KK 35 belonged to the system of IC~342 satellites, but even the authors of the
 paper \citep{kar03} themselves were not completely sure both of their measurements of the distance to KK~35 and of the actual
 belonging of KK~35 to dwarf galaxies. Nevertheless, in their subsequent publications, KK~35 remained in the lists of IC~342 satellites
 without any doubt about the nature of this object \citep{kar04}. 

We reprocessed the original HST images of KK~35. Figure~7 presents the CM diagram for the stars of KK~35 based on the photometry of the
 same HST images that were used by Karachentsev et al. (2003) in determining the distance to KK~35. Because of the method of photometry used,
 the limit of our CM diagram is slightly deeper than that of the diagram by Karachentsev et al. (2003), but, nevertheless, there is no red
 giant branch on our CM diagram (fig.~7). This means that the distance to KK~35 cannot be determined by the TRGB method based only on the
 WFPC2 images used by Karachentsev et al. (2003).

It is quite likely that these authors took the increase in the number of pseudostars appearing near the photometric limit of the images
 due to the difficulty of separating faint stars from noise fluctuations as the actual red giant branch and determined the distance on this
 basis. The presumed location of the red giant branch in the region of KK~35  that will be seen on deep images of future observations is
 marked in fig.~7.

 Figure~8 presents the $HI$ radio image of IC~342 obtained Crosthwaite et al. (2001). Whereas a strong extinction in the optical band 
reduces significantly the galaxy's sizes, making it periphery invisible, the entire spiral structure of the galaxy is seen in the radio band.
 We clearly see from fig.~8 that KK~35 is only one of several bright star-forming regions in the spiral arms of IC~342. When the apparent 
distribution of young stars is constructed, the HST WFPC2 images show that KK 35 is elongated precisely along the spiral arm of IC~342
 and the sizes of KK 35 (along its major axis) are beyond the limits of the HST image. Thus, there is no reason to consider KK~35 to be
 a separate dwarf galaxy. We believe that the object KK~35 is a fragment of the spiral arm of IC~342 with a bright star-forming region
 and must be excluded from the list of galaxies.

\section*{CONCLUSIONS} 

The distance to IC~342 that we determined ($3.93 \pm 0.10$ Mpc) differs significantly from its value assumed thus far ($3.28 \pm 0.26$ Mpc).
 This not only entails a revision of the physical parameters for IC~342 (its luminosity and mass) but also gives a completely new view on
 the structure of the IC~342/Maffei group, because some of the galaxies, KKH5, KKH 6, KKH 34, and others \citep{kar03}, that
 have previously been assumed to be located far from IC~342 are now in the zone of its influence. The spatial location of the massive
elliptical galaxy Maffei 1 is of particular importance in determining all characteristics of the IC~342/Maffei group. Whereas Maffei~1 
was initially included even in the Local Group \citep{spin71}, its distance was subsequently increased to 2 Mpc \citep{but83}
 and, in the last decade, its value varies between 3 and 4.5 Mpc \citep{lup93,dav01,dav05,fing03}. 
The most recent measurements based on the photometry of the globular clusters found give a distance of 
4.1 Mpc \citep{dav05}, which makes the two galaxies (IC~342 and Maffei 1) the unquestionable center of the group
 with a spatial separation between them of 0.8 Mpc. The Local Group of galaxies that has two massive central galaxies (M31 and our Galaxy)
 with a separation between them of approximately 0.8 Mpc has a similar structure. Naturally, less massive galaxies also belong to the 
IC~342/Maffei group. Although accurate distance determinations for them is a matter of future observations, the available results
 nevertheless indicate that the galaxies NGC 1569 \citep{gro08}, UGCA 86, and UGCA 92 belonging to the IC~342/Maffei group
 have a considerably larger distance modulus than was assumed by  Karachentsev et al. (2003) when compiling their list of members of
 the IC~342/Maffei group and calculating its dynamical characteristics. 

\newpage
\begin{table}[t]
\begin{center}
\footnotesize
\renewcommand{\tabcolsep}{4pt}
\caption{ Distance determinations for IC~ 342}
\vspace{0.2cm}
 \begin{tabular}{|clcc|} \hline
\multicolumn{1}{|c}{D, Mpc}&
\multicolumn{1}{c}{$(m-M)$}&
\multicolumn{1}{c}{Method}&
\multicolumn{1}{c|}{Authors}\\ \hline
  1.5  & 25.9   & H II regions    & \citep{abl71}                \\
  7.9  & 29.5   & H II regions    & \citep{san74}                 \\
  2.9  & 27.34  & H II regions    & \citep{vau78}                 \\
2.08   & 26.60  & Brightest stars & \citep{kar93}                  \\
3.28   & 27.58  & Cepheids        & \citep{sah02}                  \\
 3.93  & 27.97  &  TRGB           & This paper               \\
\hline
\end{tabular}
\end{center}
\end{table}

\begin{figure}[ht]
\centerline{\includegraphics[angle=0, width=14cm]{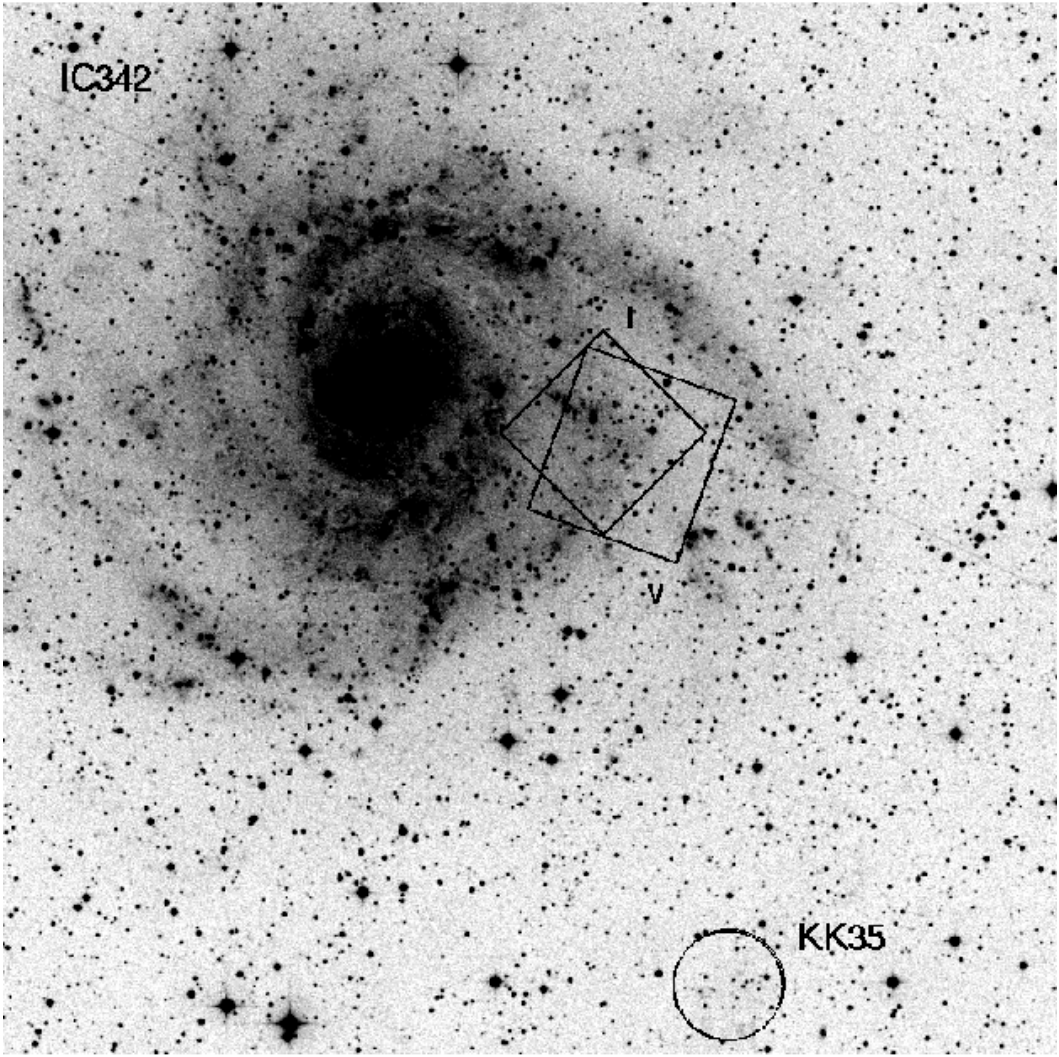}}
\caption{DSS2 R-band image of IC~342. The two ID10579 and ID10768 image fields whose intersection was used for our stellar
 photometry are marked. The position of KK 35, a presumed dwarf galaxy from the list by Karachentsev et al. (2003),
 is marked. The size of the DSS2 image is $25\arcmin \times 25\arcmin$ ; the north is at the top.}
\end{figure}

\begin{figure}[ht]
\centerline{\includegraphics[angle=0, width=14cm]{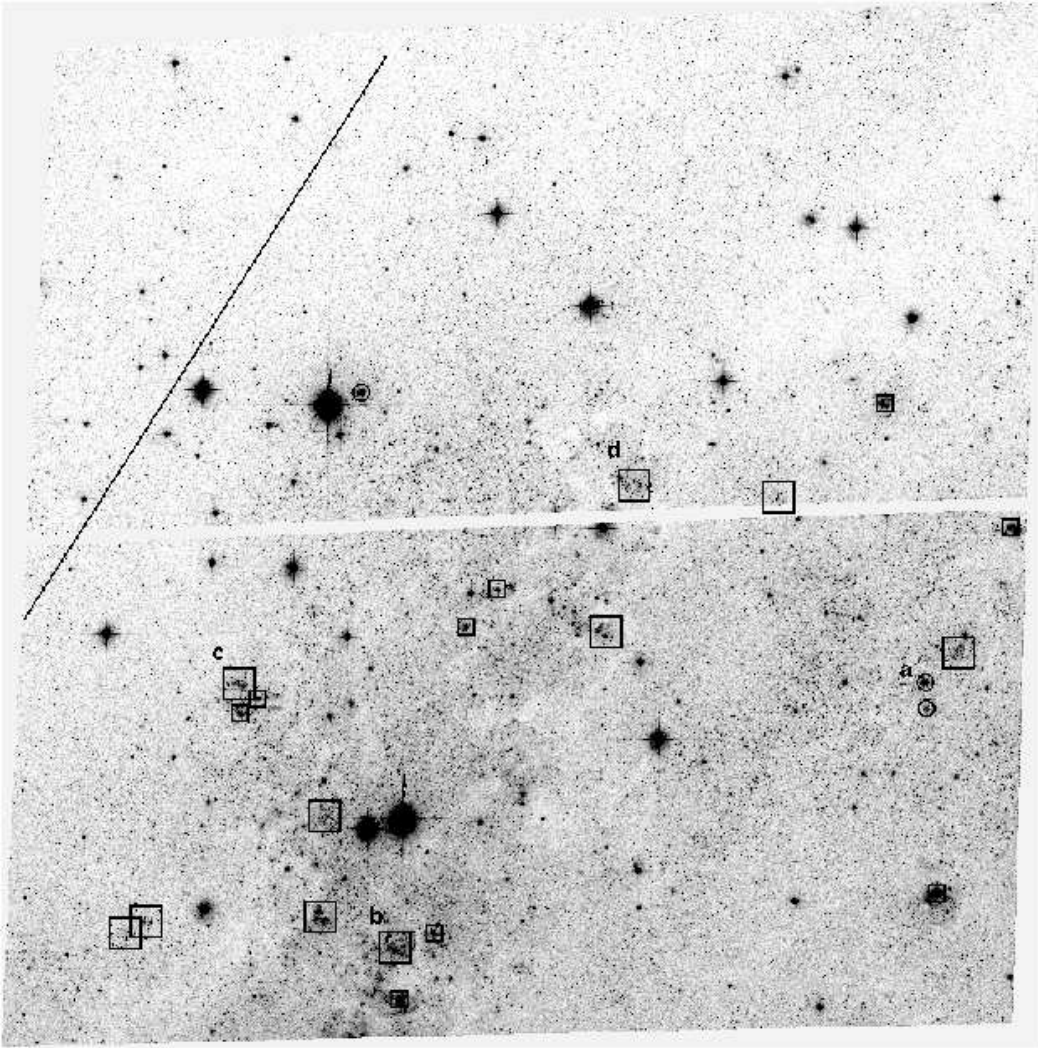}}%
\caption{F606W-filter (V) image of the IC~342 field used to search for star clusters and to determine the distance by the
 TRGB method. The straight line serves as the boundary of the region (the upper left corner of the field) where the red
 giants were selected to determine the distance (see fig.~4 below). The circles, small squares, and large squares denote 
globular clusters, compact clusters, and star complexes, respectively. The letters a, b, c, and d denote the objects whose
 CM diagrams are presented in fig.~6.}
 \end{figure}

\begin{figure}[ht]
\centerline{\includegraphics[angle=0, width=10cm, bb=203 325 429 560,clip]{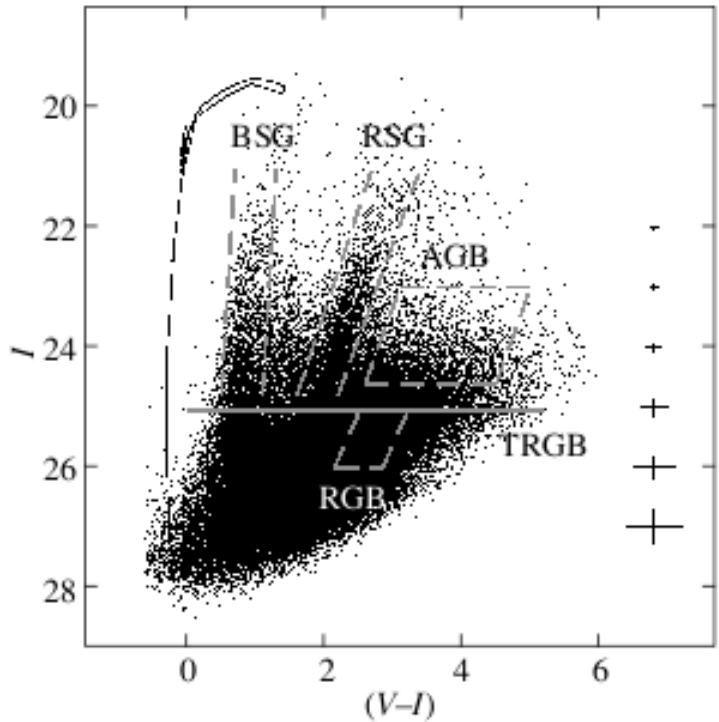}}
\caption{Hertzsprung-Russell diagram for the stars of the region of intersection between the two HST ACS/WFC images (see fig.~1).
 The positions of the main types of stars are marked: blue supergiants (BSG), red supergiants (RSG), intermediate-age stars (AGB),
 and red giants (RGB). The stellar isochrone with an age of 8 Myr and metallicity Z = 0.02 marks the position that the blue
 supergiant branch would occupy in the absence of extinction. The horizontal line marks the location of the upper RGB boundary
 (TRGB jump). The mean errors of our stellar photometry are shown to the right on the diagram.}
\end{figure}

\begin{figure}[ht]
\centerline{\includegraphics[angle=270, width=10cm, bb=130 136 444 450,clip]{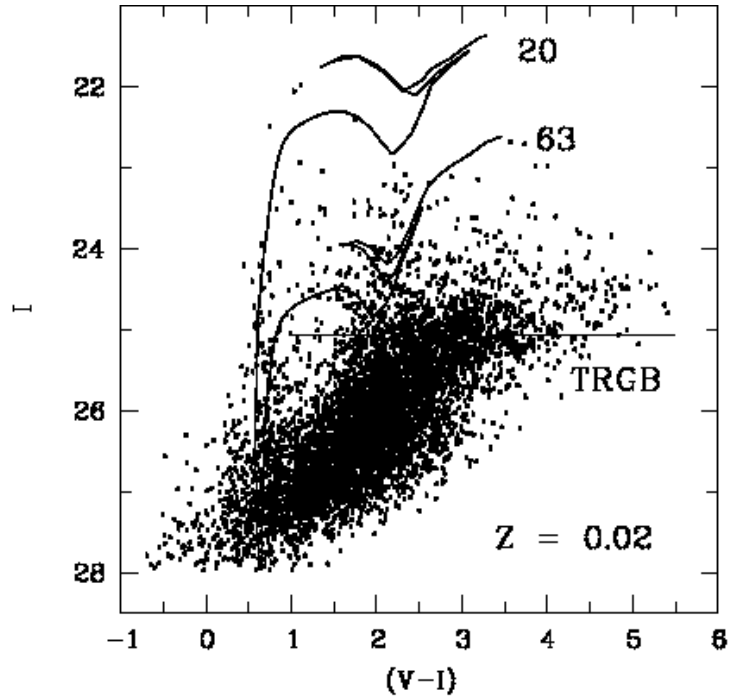}}%
\caption{Hertzsprung-Russell diagram for the stars of the field (see fig.~2) outside the regions of intense star formation. 
The boundary of the field is marked by the straight line in fig.~2. The TRGB boundary at I = 25.06 is marked on the diagram. 
The isochrones pointing to the presence of a few young high-metallicity stars among the photometered stars were constructed
 for ages of 20 and 65 Myr and metallicity Z = 0.02. The reddening $E(V -I) = 0.82\pm 0.02$ 
 for this field was determined by
 choosing optimal isochrones for this CM diagram.}
\end{figure}

\begin{figure}[ht]
\centerline{\includegraphics[angle=270, width=12cm, bb=130 136 454 460,clip]{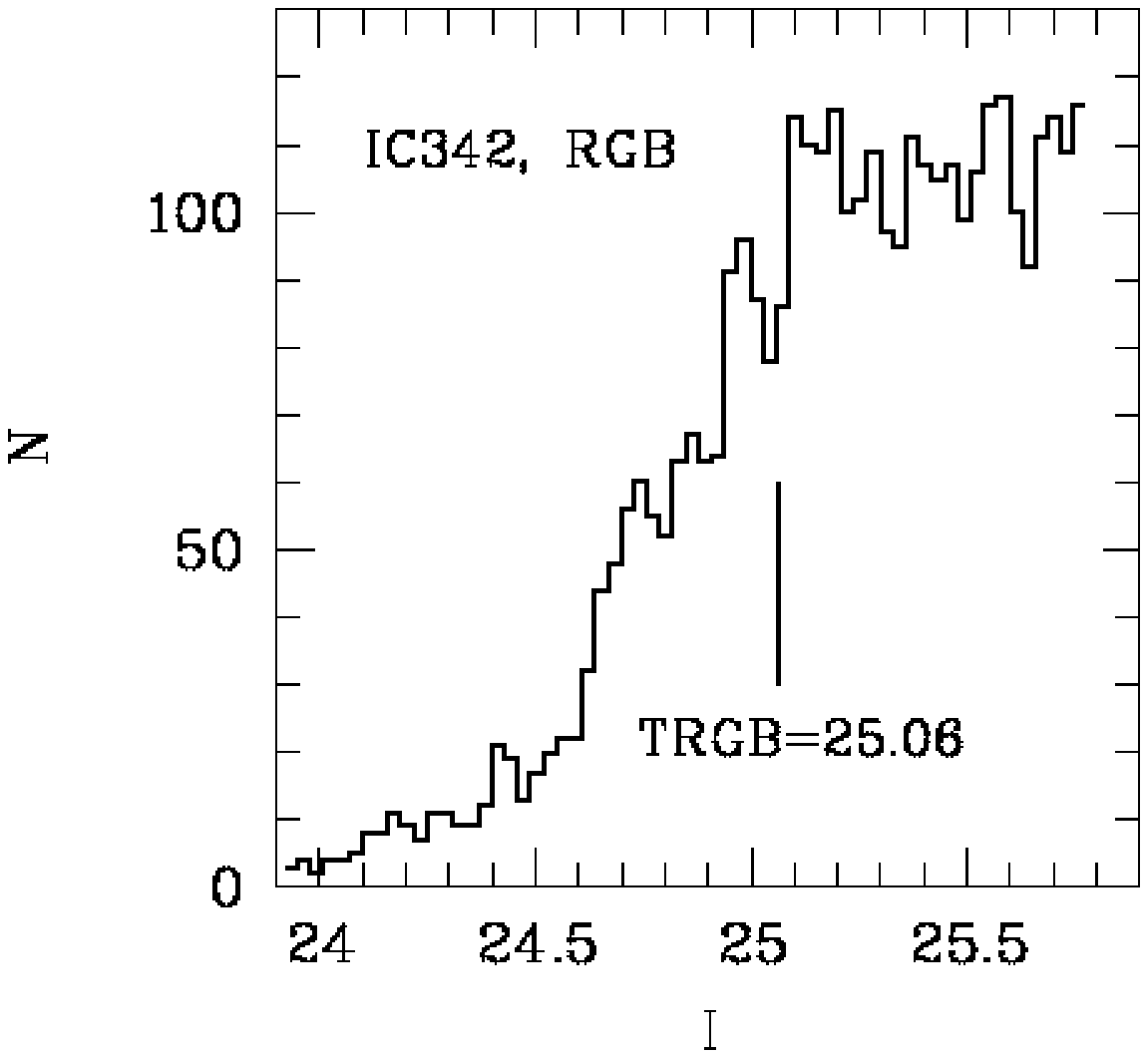}}%
\caption{ Luminosity function for the red giants of the diagram in fig.~4. The beginning of the red giant branch (TRGB jump)
 is observed at $I = 25.06$, which corresponds to the distance $D = 3.93$ Mpc corrected for the reddening $E(V -I) = 0.82$ found.}
\end{figure}

\begin{figure}[ht]
\centerline{\includegraphics[angle=270, width=16cm, bb=33 68 523 596,clip]{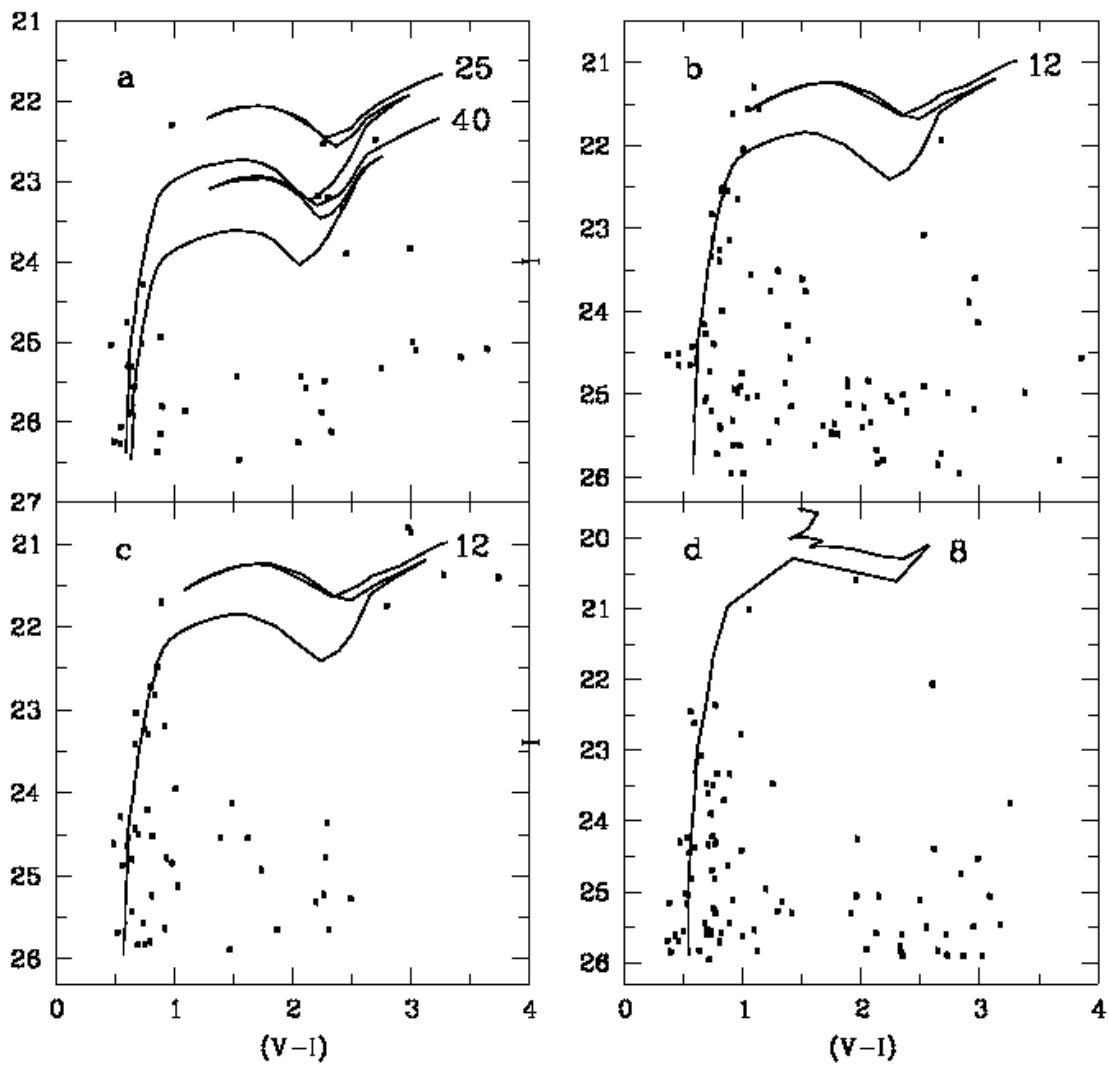}}%
\caption{Hertzsprung-Russell diagrams for the stars of a young globular cluster (a) and three small star complexes (b, c, d). 
Optimal isochrones from \citep{ber94} were found for each object. The reddening for the four objects varies within the range 
$0.78 < E(V -I ) < 0.83$. The metallicity of young stars equal to the solar metallicity (Z = 0.02) was found for the cluster (a) 
with a sufficient number of red supergiants. The metallicity of red supergiants in the remaining objects is consistent with this value.}
\end{figure}

\begin{figure}[ht]
\centerline{\includegraphics[angle=270, width=10cm, bb=130 136 454 460,clip]{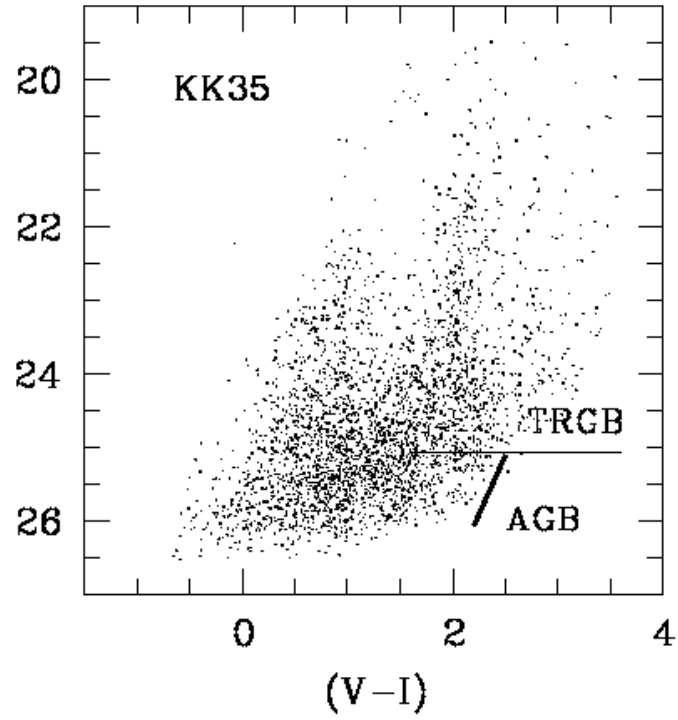}}%
\caption{Hertzsprung-Russell diagram for the stars of KK 35 based on the photometry of WFPC2 images.
 There are reddened stars of the blue and red supergiant branches, but there is no red giant branch beyond the photometric limit 
of the images used. The oblique and horizontal lines mark the presumed location of the upper part of the red giant branch (RGB) and 
the boundary of this branch (TRGB jump), respectively.}
\end{figure}

 \begin{figure}[ht]
 \centerline{\includegraphics[angle=0, width=17cm]{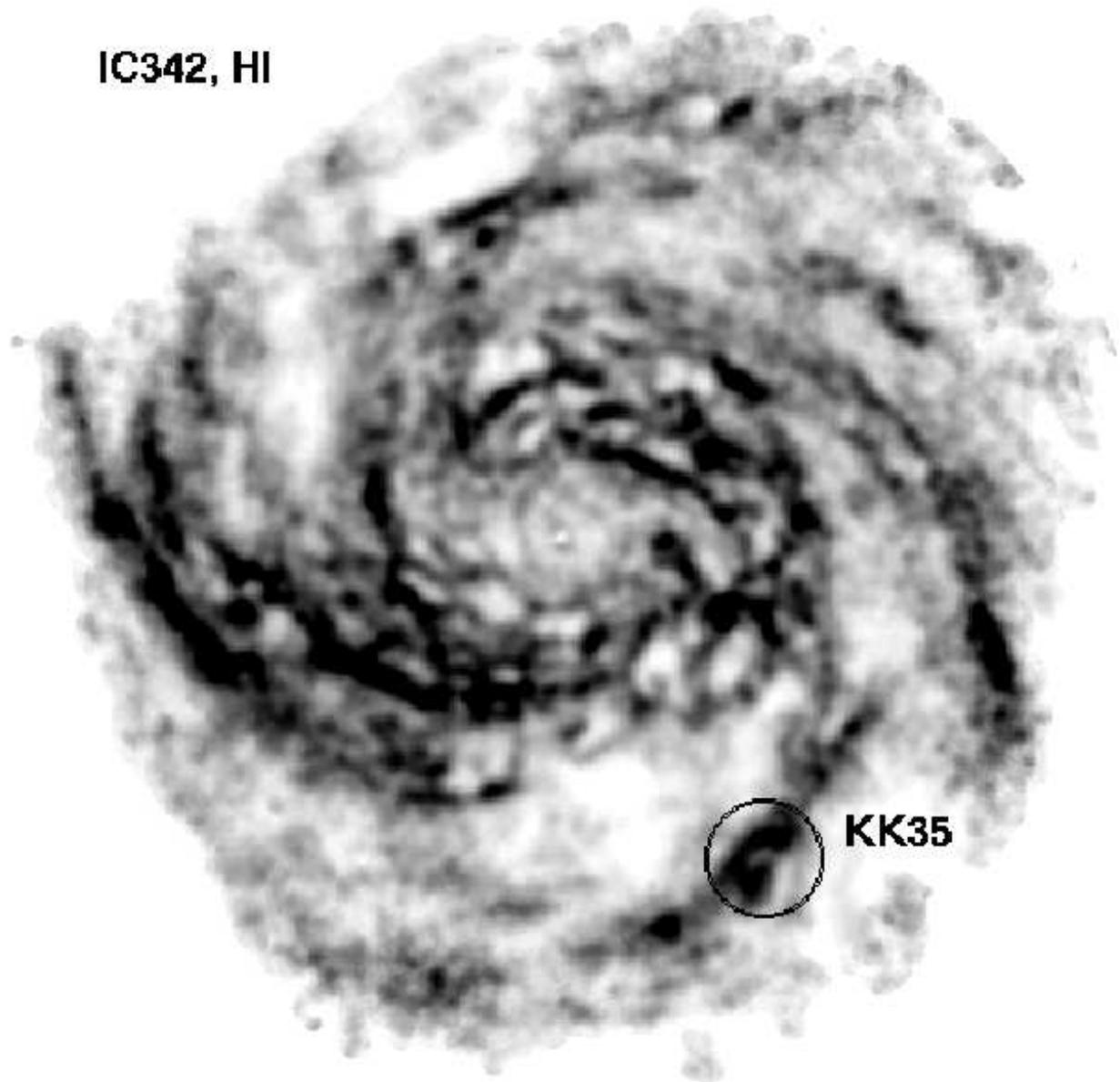}}%
\caption{ $HI$ radio image of IC~342. Extended spiral arms are seen and the bright part of one of them coincides with the "galaxy" KK 35.}
 \end{figure}


\newpage
\begin{thebibliography}{}
{\small
\bibitem[Ables (1971)]{abl71}Ables, H. D., Publ. U.S. Naval Observ. \ 20, 1 (1971). 
\bibitem[Aparicio (1994)]{apa94} Aparicio, A., \apj \ 437, L27 (1994).
\bibitem[Aparicio et al. (2000)]{apa00} Aparicio, A., Tikhonov, N. A., \&  Karachentsev, I. D., \aj \ 119, 177 (2000). 
\bibitem[Bertelli et al. (1994)]{ber94} Bertelli, G., Bressan, A., Chiosi, C. et al., \aap \ 106, 275 (1994). 
\bibitem[Burstein \& Heiles (1982)]{bur82} Burstein, D., \& Heiles, C., \aj \ 87, 116 (1982). 
\bibitem[Buta \& McCall (1983)]{but83} Buta, R.J., \& McCall, M.L., \mnras \ 205, 131 (1983). 
\bibitem[Buta \& McCall (1999)]{but99} Buta, R.J., \& McCall, M.L., \apjs  \ 124, 33 (1999). 
\bibitem[Crosthwaite et al. (2001)]{cro01} Crosthwaite, L.P., Turner, J.L. , Hurt, R.L.  et al., \aj \ 122, 797 (2001). 
\bibitem[Davidge \& van den Bergh (2001)]{dav01} Davidge, T.J., \& van den Bergh, S., \apj \ 553, L133 (2001). 
\bibitem[Davidge \&  van den Bergh (2005)]{dav05} Davidge, T.J., \&  van den Bergh S., \pasp \ 117, 589 (2005). 
\bibitem[Fingerhut et al.(2003)]{fing03} Fingerhut, R.L., McCall, M.L., De Robertis, M. et al., \apj \ 587, 672 (2003).
\bibitem[Grocholski et al.(2008)]{gro08} Grocholski, A.,  Aloisi, A., van der Marel, R. et al., \apj  \ 686, L79 (2008). 
\bibitem[Hoessel et al. (1990)]{hoe90} Hoessel, J.G., Abbott, M. J.,  Saha, A. et al., \aj \ 100, 1151 (1990). 
\bibitem[Hoessel et al. (1994)]{hoe94} Hoessel, J.G.,  Saha, A.,  Krist, J., \&  Danielson G. E., \aj \ 108, 645 (1994). 
\bibitem[Hoessel et al. (1998)]{hoe98} Hoessel, J.G., Saha, A., \&   Danielson, G.E., \aj \ 116, 1679 (1998).
\bibitem[Huchtmeier et al. (1997)]{huch97} Huchtmeier, W.K.,  Karachentsev, I.D., \&  Karachentseva V. E., \aap \ 322, 375 (1997). 
\bibitem[Karachentsev et al. (2004)]{kar04} Karachentsev, I.D.,  Karachentseva, V.E.,  Huchtmeier, W.K., \& Makarov, D.I., \aj \ 127, 2031 (2004). 
\bibitem[Karachentsev et al. (2003)]{kar03} Karachentsev, I.D.,  Sharina M. E.,  Dolphin A.E.,  \& Grebel, E.K., \aap \ 408, 111 (2003). 
\bibitem[Karachentsev \& Tikhonov (1993)]{kar93}  Karachentsev, I.D., \&  Tikhonov, N.A., \aaps \ 100, 227 (1993). 
\bibitem[Krismer et al. (1995)]{kris95}  Krismer, O.M., Tully B., \& Gioia, I., \aj \ 110, 1584 (1995). 
\bibitem[Lee et al. (1993)]{lee93} Lee, M.G.,  Freedman, W.L., \&  Madore, B.F., \apj \ 417, 553 (1993). 
\bibitem[Luppino \&  Tonry (1993)]{lup93} Luppino, G.A. \&  Tonry, J.L., \apj \ 410, 81 (1993). 
\bibitem[Makarov et al. (2006)]{mak06} Makarov, D.,  Makarova, L. , Rizzi, L. et al., \aj \ 132, 2729 (2006). 
\bibitem[McCall (1989)]{mcc89} McCall, M.L., \aj \ 97, 1341 (1989). 
\bibitem[Rizzi et al. (2007)]{riz07} Rizzi, L., Tully, B.,  Makarov, D. et al., \apj \ 661, 815 (2007). 
\bibitem[Saha et al. (2002)]{sah02} Saha, A. , Claver, J., \&  Hoessel, J.G., \aj \ 124, 839 (2002). 
\bibitem[Sandage \& Tammann (1974)]{san74} Sandage, A., \&  Tammann, G.A.,  \apj \ 194, 559 (1974). 
\bibitem[Schlegel et al. (1998)]{sch98} Schlegel, D.J., Finkbeiner, D.P., \&  Davis M. , \apj \ 500, 525 (1998). 
\bibitem[Spinrad (1971)]{spin71}  Spinrad, H., Sargent, W.L.W.,  Oke J.B. et al., \apj \ 163, 25 (1971). 
\bibitem[Stetson (1994)]{stet94} Stetson,  P.B., \pasp \ 106, 250 (1994).
\bibitem[Tolstoy et al. (2007)]{tol07}  Tolstoy, E. , Gallagher, J.S.,  Cole, A.A.,  \& Richer, M.G., \apj \ 655, 814 (2007). 
\bibitem[de Vaucouleurs (1978)]{vau78}  de Vaucouleurs, G., \apj \ 224, 710 (1978).}
\end{thebibliography}
\end{document}